 \documentclass{emulateapj}

\usepackage{txfonts}
\usepackage{psfrag}
\usepackage{color}
\usepackage{graphicx}
\usepackage{times}
\usepackage{natbib}
\bibpunct{(}{)}{;}{a}{}{,} 
\newcommand{\kms}{{\rm\,km\,s^{-1}}\,}

\newcommand{\mh}{{\rm\, [M/H]}}

\def\gsim { \lower .75ex \hbox{$\sim$} \llap{\raise .27ex \hbox{$>$}} }
\def\lsim { \lower .75ex \hbox{$\sim$} \llap{\raise .27ex \hbox{$<$}} }
\newcommand{\pc}{\rm\,pc}
\newcommand{\kpc}{{\rm\,kpc}}
\newcommand{\dex}{{\rm\,dex}}

\newcommand{\Ro}{R_0}

\newcommand{\vo}{V_0}

\newcommand{\vphi}{v_{\phi}}
\newcommand{\vrr}{v_{R}}

\newcommand{\Vs}{V_\odot}
\newcommand{\Us}{U_\odot}
\newcommand{\Ws}{W_\odot}

\newcommand{\kmskpc}{\mbox{\rm\,km\,s$^{-1}$\,kpc$^{-1}$}}

\newcommand{\cutt}[1]{}

\def\aap{A\&A}

\def\aj{AJ}

\def\2s{2-$\sigma$}
\def\3s{3-$\sigma$}

\shorttitle{Imprints of the bar on the thick disk}
\shortauthors{Antoja et al.}

\begin{document}


\title{The imprints of the Galactic bar on the thick disk with RAVE}


\author{T. Antoja$^{1,2, \star}$, 
 G. Monari$^{2,3}$,
 A. Helmi$^{2}$,
  O. Bienaym\'e$^{3}$,
  J. Bland-Hawthorn$^{4}$,
  B. Famaey$^{3}$,
   B. K. Gibson$^{5}$,
  E.~K.\ Grebel$^{6}$,
  G. Kordopatis$^{7}$,
   U. Munari$^{8}$,
  J. Navarro$^{9}$,
  Q. Parker$^{10,11,12}$,
  W.A. Reid$^{11}$,
  G. Seabroke$^{13}$,
  M. Steinmetz$^{7}$,
 T. Zwitter$^{14}$ 
}        

 \affil{$^1$ Scientific Support Office, Directorate of Science and Robotic Exploration, European Space
Research and Technology Centre (ESA/ESTEC), Keplerlaan 1, 2201 AZ Noordwijk, The
Netherlands; tantoja@cosmos.esa.int,
   \\$^2$ Kapteyn Astronomical Institute, University of Groningen, PO Box 800, 9700 AV Groningen, the Netherlands,
 \\$^3$   Observatoire astronomique de Strasbourg, Universit\'e de Strasbourg, CNRS,  UMR 7550, 11 rue de l'Universit\'e F-67000 Strasbourg, France,
   \\$^4$  Sydney Institute for Astronomy, School of Physics A28, University of Sydney, NSW 2006, Australia,
 \\$^5$   Jeremiah Horrocks Institute, University of Central Lancashire, Preston, PR1 2HE, United Kingdom,
 \\$^6$  Astronomisches Rechen-Institut, Zentrum f\"ur Astronomie der Universit\"at Heidelberg, M\"onchhofstr.\ 12--14, 69120 Heidelberg, Germany,
  \\$^7$  Leibniz-Institut f\"ur Astrophysik Potsdam (AIP), An der Sternwarte 16, D - 14482, Potsdam, Germany,  
 \\$^8$   INAF Osservatorio Astronomico di Padova, 36012 Asiago (VI), Italy, 
  \\$^9$  University of Victoria, Victoria BC, Canada V8P 5C2,
 \\$^{10}$  Research Centre in Astronomy, Astrophysics and Astrophotonics, Macquarie University, Sydney, NSW 2109 Australia,
  \\$^{11}$  Department of Physics and Astronomy, Macquarie University, Sydney, NSW, 2109, Australian,
  \\$^{12}$  Australian Astronomical Observatory, PO Box 915, North Ryde, NSW 1670, Australia,
  \\$^{13}$  Mullard Space Science Laboratory, University College London, Holmbury St Mary, Dorking, RH5 6NT, UK,
\\$^{14}$  University of Ljubljana, Faculty of Mathematics and Physics, Jadranska 19, 1000 Ljubljana, Slovenia
}
\altaffiltext{$^{\star}$}{ESA Research Fellow.}


\begin{abstract}
We study the kinematics of a local sample of stars, located within a cylinder of $500\pc$ radius centered on the Sun,  in
     the RAVE dataset. We find clear asymmetries in the $\vrr$-$\vphi$ velocity
     distributions of thin and thick disk stars:  there are more stars moving radially outwards for low azimuthal velocities 
    and more radially inwards for high azimuthal velocities.
 Such asymmetries have
     been previously reported for the thin disk as being due to the
     Galactic bar, but this is the first time that the same type of
     structures are seen in the thick disk. Our findings imply that the
     velocities of thick disk stars should no longer be
     described by Schwarzschild's, multivariate Gaussian or purely axisymmetric distributions. Furthermore, the nature of previously reported
     substructures in the thick disk needs to be revisited as these
     could be associated with dynamical resonances rather than to accretion
     events. It is clear that dynamical models of the Galaxy must
     fit the 3D velocity distributions of the disks, rather than the projected 1D,
     if we are to understand the Galaxy fully. 
\end{abstract}





\section{Introduction}
\label{sec:intro}

The velocity distribution of stars in the Solar neighborhood contains
key information about the current dynamical state of our Galaxy and
also about its history. The kinematics of stars can be used to derive
both the mass distribution of the Milky Way through sophisticated
dynamical models, as well as to identify accretion events.

The velocities of thin disk stars are often described by the
Schwarzschild distribution function, which considers each of the
velocity components separately. However, data from the Hipparcos
mission (\citealt{Perryman97}; ESA 1997) and later from the Geneva-Copenhagen survey \citep{Holmberg09}
have revealed a
more complex distribution with significant overdensities and structure
\citep{Dehnen98,Famaey05,Antoja08}. Some of the over-densities
and distortions of the velocity distribution appear to be the
imprints of the non-axisymmetric components of the Milky Way, namely
the spiral arms and Galactic bar
(e.g. \citealt{Dehnen00,DeSimone04,Sellwood10,Antoja11,McMillan11}). Streams or moving groups are
formed by stars on orbits that are close to resonant with the natural frequencies of the
spiral arms and/or the bar. Examples of 
such moving groups that are heterogeneous in age and chemical composition are
 the Pleiades, Hyades, Sirius,
and Hercules streams (the latter very likely driven by the bar).

On the other hand, the velocity distribution of the thick disk has
been studied in less detail thus far because of limitations in the
size of volume complete samples. In practice, this has implied that
Gaussian distributions were used to fit the kinematics
of thick disk stars
\citep[see][for a recent discussion on how Gaussian functions poorly fit all velocity components]{Binney14,Sharma14}.
Further, substructures have also been
reported in the thick disk
\citep[e.g.][]{Gilmore02,Navarro04,Helmi06}, 
identified through statistical comparisons with Galactic models or
with simple kinematic models such as those discussed above.

Many of these substructures have been attributed to accretion events,
as these typically leave behind streams of stars with similar
velocities that do not necessarily appear to be spatially coherent
near the Sun because of the short mixing timescales in the inner
Galaxy. However, recent modeling has shown that the impact of spiral
arms \citep{Solway12,Faure14} and the Galactic bar on the kinematics of stars in the thick disk
is non-negligible. For example, 
\citet{Monari13} and \citet{Monari14b} have found in their simulations that there is as much
resonant trapping in the  thick disk as in the thin disk. 
Another clear signature of the impact of the bar in their thick disk simulations
is a bimodality in the velocity distribution for stars
located near the Outer Lindblad Resonance, similar to that observed in the thin disk. 


Here we explore whether these features are
present in local samples of thick disk stars, especially now that such
samples have increased in size by large factors, (as in e.g. LAMOST and SEGUE, \citealt{Cui12} and \citealt{Yanny2009}, respectively). For example 
the Geneva-Copenhagen survey contained $\sim17,000$ stars, while  $\sim400,000$ stars have now been measured by the RAdial Velocity Experiment (RAVE) \citep{Steinmetz06}. We report on the analysis of the
{\it local}  RAVE dataset, and find indeed clear asymmetries/structures in the
velocity distributions of local thick disk stars, which can be attributed to
the resonant interaction with the Galactic bar. In Section
\ref{sec:data} we present the dataset and the selection of the different populations, in Section \ref{sec:stats} the analysis,
and we conclude in Section \ref{sec:concl} with a discussion on the
implications of our findings.

\section{Observations and data selection}
\label{sec:data}

In this study we use the RAVE Data Release 4 (DR4) \citep{Kordopatis13}. The stellar atmospheric parameters of the
DR4 are computed using two different pipelines, presented by  \cite{Kordopatis11} and
\citet{Siebert11b}. The
stellar parallaxes that we use were obtained through the Bayesian distance-finding
method of \citet{Binney14}.

 First we select stars with i)
signal-to-noise ratio better than 20, ii) the first morphological flag indicating that they are normal stars \citep{Matijevic12}, and iii) converged algorithm of computation of the physical parameters. From these, we further select those in a cylinder with radius of $500\pc$ centred on the Sun's position. 
This results in a sample of 162153 stars with
6D phase-space information, of which $76\%$ are dwarf stars and $24\%$ are giants.
The DR4 proper motions were
compiled from several catalogs and here we use UCAC4 \citep{Zacharias13}.


Following \citet{Reid14} we assume that the Sun is at $X =-8.34\kpc$ and
take a circular velocity at the Sun of $\vo=240\kms$. For the
velocity of the Sun with respect to the Local Standard of Rest we adopt
$(\Us, \Vs, \Ws)=(10,12,7)\kms$ \citep{Schonrich10}. The
resulting value of $(\vo+\Vs)/\Ro$ is $30.2\kmskpc$, which is
compatible with that from the reflex motion of the Sgr A*
$30.2 \pm 0.2 \kmskpc$ \citep{Reid04}. With these values, we compute
the stars' cylindrical velocities: $\vrr$
(positive towards the Galactic center, in consonance with the usual $U$
velocity component) and $\vphi$ (towards the direction of rotation).

 \begin{table*}
 \setlength{\tabcolsep}{1.5pt}
 \centering 
 \caption{Properties of the different cuts.}  \label{t:samples}   

 \begin{tabular}{l|ll|rrrrrrr|rrr|rr}      
 \hline\hline                 
 &    &    & \multicolumn{1}{c}{N}& \multicolumn{1}{c}{$e_Z$} & \multicolumn{1}{c}{$e_{dist_{2d}}$} & \multicolumn{1}{c}{$e_{\vrr}$}& \multicolumn{1}{c}{$e_{\vphi}$}& \multicolumn{1}{c}{$e_{{v_Z}}$}& \multicolumn{1}{c}{$e_\mh$}& \multicolumn{3}{|c|}{RAVE-fit} & \multicolumn{2}{c}{fit 2} \\
 &    &    & \multicolumn{1}{c}{}& \multicolumn{1}{c}{$\kpc$} & \multicolumn{1}{c}{$\kpc$} & \multicolumn{1}{c}{$\kms$}& \multicolumn{1}{c}{$\kms$} & \multicolumn{1}{c}{$\kms$}& \multicolumn{1}{c}{$\dex$}& \multicolumn{1}{|c}{thin}& \multicolumn{1}{c}{thick} & \multicolumn{1}{c|}{halo} & \multicolumn{1}{c}{thin} & \multicolumn{1}{c}{thick} \\\hline 
 1& $\mh \ge -0.1 \dex$ & $|Z| \le 0.5 \kpc$& 47883& 0.05& 0.05&  5.0&  4.3&  3.3& 0.10& 96  &4    &0.5 &99.8&0.2\\
 2& $\mh < -0.45 \dex$  & $|Z| > 0.5 \kpc$  &  5123& 0.38& 0.13& 15.7& 15.1&  5.2& 0.10&0.2  &88   &11  &13  &87\\
 3& $\mh < -0.7 \dex$   & $|Z| \le 0.5 \kpc$& 21624& 0.06& 0.06&  6.6&  5.0&  4.0& 0.12&0.7  &78   &22  &3   &97\\
 4& $\mh < -0.7 \dex$   & $|Z| > 0.5 \kpc$  &  2939& 0.40& 0.13& 17.3& 18.0&  5.4& 0.10&0.003&68   &32  &0.7 &99.3\\
 \hline 
 \end{tabular}
 \tablecomments{The first columns show the cuts, the number of stars N, median errors in vertical position $Z$, in horizontal distance in the plane from the Sun $dist_{2d}$, in the velocity components ($\vrr$, $\vphi$ and $v_Z$), and in the metallicity $\mh$. The last 5 columns show the thin disk, thick disk and halo fraction of stars computed according to different models (named RAVE-fit and fit 2, see text for caveats with respect to the estimated halo fraction).}\end{table*}

From the selected sample we consider 4 different subsets of stars
based on their height and their metallicity to maximize or minimize the
number of thin or thick disk stars. The properties of each
subset and relative thin/thick/halo fractions are listed in
Table~\ref{t:samples}. Two of the subsets are located on the
plane but have metallicities corresponding to thin (1) and to thick
(3) disk components, respectively. The other two are located far from
the plane and have intermediate (2) and low (4) metallicities and
could be both associated with the thick disk. 

For each subset it is important to estimate the fraction of stars that
could belong to a different population than desired. We have derived
two different estimates of these fractions for each of the samples. The first estimate, which
we term RAVE-fit, is based on 
 an admittedly simplistic three Gaussian 
population model (old
thin disk, thick disk and halo) fit to the metallicity distribution to
RAVE data by \citet[][their tables 1 and 2]{Kordopatis13b}. We use the
fits derived for {\it all} stars with galactocentric radius between
$7.5\kpc$ and $8.5\kpc$, to estimate the population fraction for
samples (1) and (3). Since these samples have an additional constraint, namely
$|Z|\leq0.5\kpc$, the fractions of thick disk and
halo stars are probably overestimated. For samples (2) and (4) we use the RAVE fits derived for
stars $1<|Z|<2\kpc$ in the same radial range. In this case, since in our samples we consider all
stars with $| Z| >0.5\kpc$, it is likely that the fraction of thin and thick
disk stars is underestimated, while that of the halo is overestimated. In fact, if we assume that the halo has no net rotation and that all stars with $\vphi<0$ belong to the halo,
we can estimate the fraction of halo stars as twice that of stars with $\vphi<0$.
We find this to be of only 3\%, 0.7\% and 4\% for samples (2), (3) and (4), respectively, 
i.e. much smaller than the fractions obtained through the RAVE-fit.

The second estimate (fit 2) of the contamination in our subsets is
based on a simple model with two populations (thin and thick disk)
with specified density and metallicity distributions. We use
two exponential disks with vertical scale heights of
$h_{z,thin}=0.3\kpc$ and $h_{z,thick}=0.9\kpc$ and scale lengths of
$h_{R,thin}=2.6\kpc$ and $h_{R,thick}=3.6\kpc$
\footnote{The scale lenghts and density normalization of the disks are uncertain, see e.g. \citealt{Bensby11}).}, and a local
density normalization of $12\%$, all as measured by \citet{Juric08}\footnote{\citet{Robin03} gives a normalization of $~27\%$ for the intermediate-age to old thin to thick disk. This would yield an even lower thin disk contamination in samples (2) to (4).}. 
We also assume Gaussian metallicity distributions with means
$\langle\mh_{thin}\rangle=-0.1$ and $\langle\mh_{thick}\rangle=-0.78$ and dispersions
$\sigma_{{\mh}_{thin}}=0.2$ and $\sigma_{{\mh_{thick}}}=0.3$ (similar
to the intermediate-old thin and thick disk
populations of \citealt{Robin03}, respectively). We estimate the fraction of each population by integrating between 
the given ranges of metallicities and heights. For samples (2) and (4) we assume a maximum height of $1.5\kpc$.

The population fractions estimated with the
two methods (RAVE-fit and fit 2) indicate that the contamination of
thick disk stars in sample (1) is very low. On the other hand, samples (2), (3) and (4) are clearly dominated by the thick disk as desired.

\section{Statistical analysis}
\label{sec:stats}

\begin{figure*}
\centering
\includegraphics[width=0.6\textwidth]{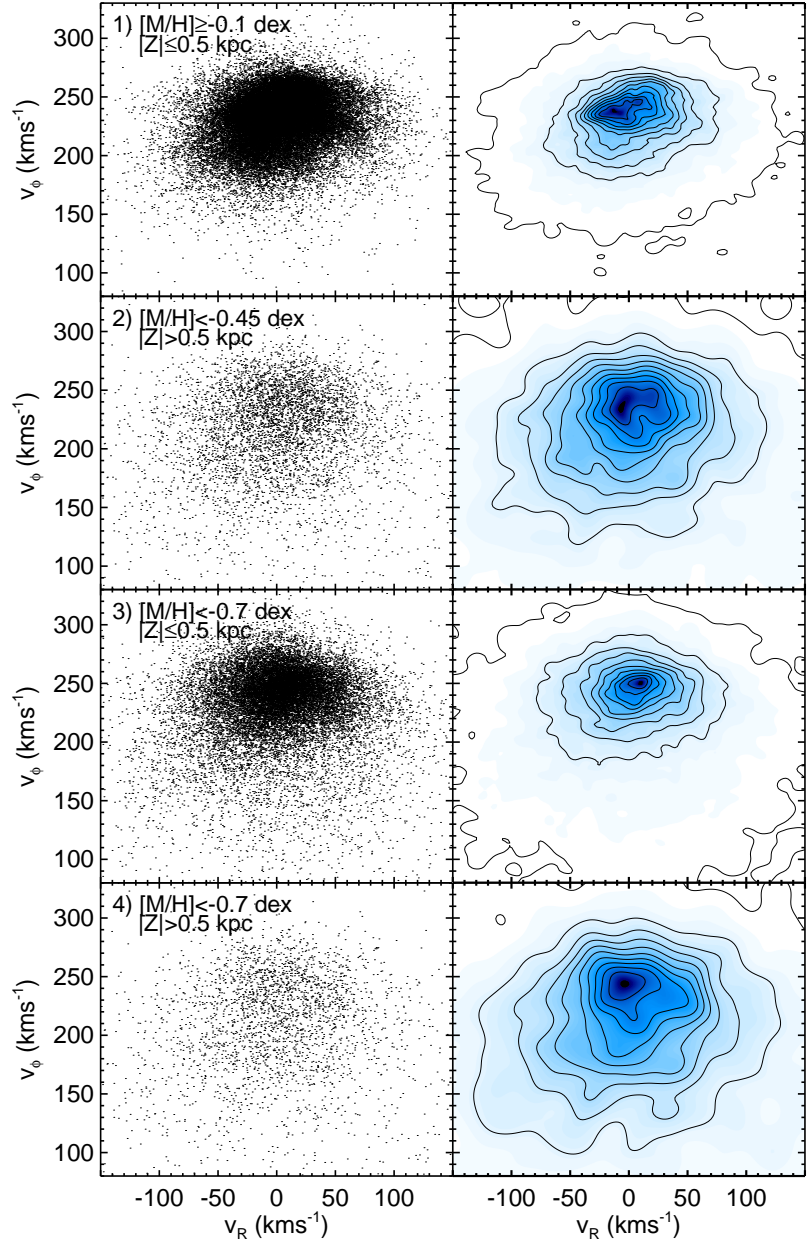}
\caption{Velocities of stars in the different cuts  of the RAVE dataset of Table~\ref{t:samples}. Left column: scatter plots. Right column: density obtained through the Epanechnikov adaptive kernel density estimator method \citep{Silverman86} with an adaptability exponent of 0.1. The density was estimated in a uniform grid of $2\kms$. The black contours indicate the following levels in units of the maximum density: 0.005, 0.1, 0.2, 0.3, 0.4, 0.5, 0.6, 0.7, 0.8, 0.9, 0.995.}
\label{fig:velocity-rave2}
\end{figure*}

In Fig.~\ref{fig:velocity-rave2} we show the velocity distributions of
the different samples using scatter plots (left) and a kernel
density estimator (right, see caption for details).  The velocity distribution of the thin disk, subset (1), is not homogeneous and depicts overdensities and
asymmetries, as already reported  in \citet{Antoja12}  for RAVE thin disk stars. We see a clear asymmetry: stars
with $\vphi\lesssim220 \kms$ are shifted to the left part of the
distribution ($\vrr<0$). Interestingly, this 
asymmetry is visible in the thick
disk subsets for both scatter and density plots shown in the remaining panels.

\begin{figure*} 
\centering
\includegraphics[height=0.6\textheight]{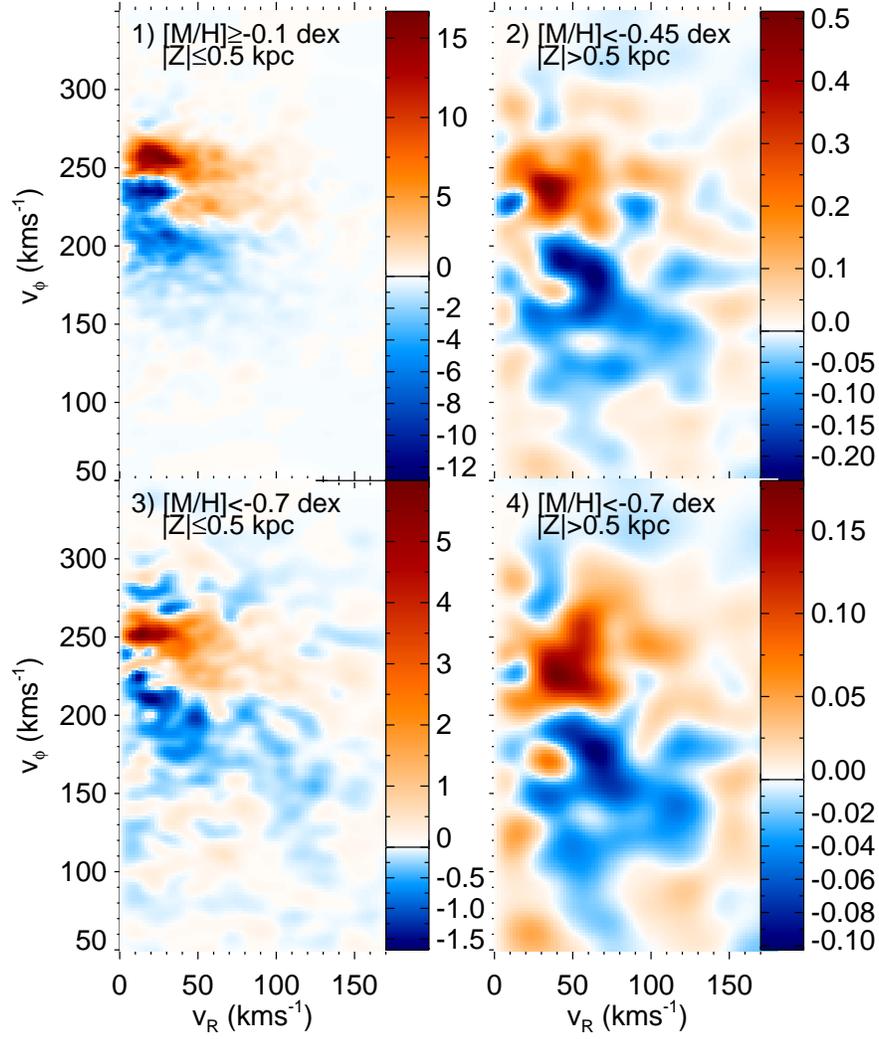}
\caption{Differences in the density field (depicted in the right panels of Fig.~\ref{fig:velocity-rave2}) for stars with positive and negative $\vrr$ for the different subsets. The density was normalized to the number of stars in each sample and, therefore, the color bars indicate difference in the number of stars in each pixel of the grid (of $2\kms$).}
\label{fig:dif2d}
\end{figure*} 

\begin{figure*} 
\centering
\includegraphics[height=0.5\textheight]{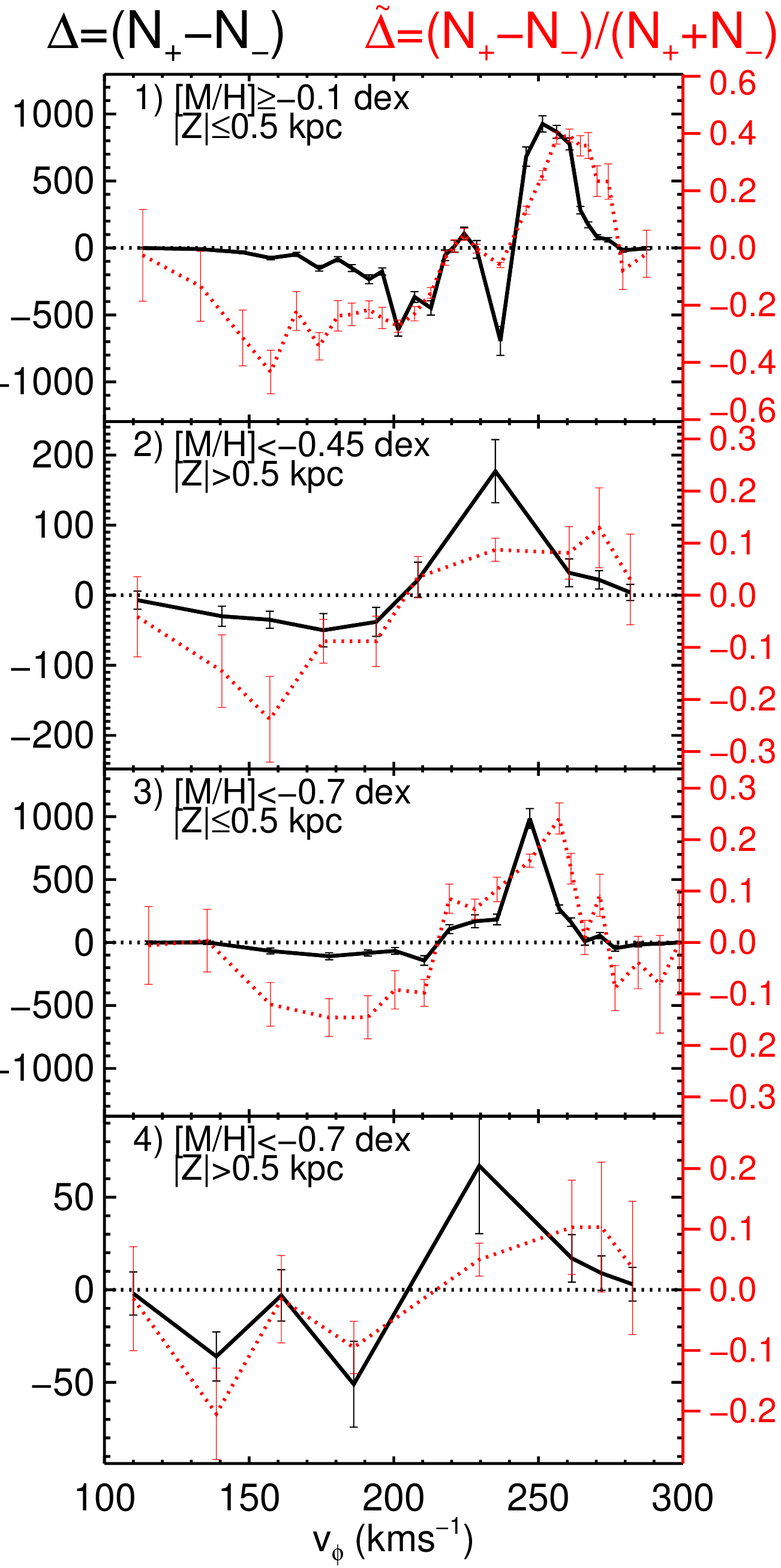}
\includegraphics[height=0.5\textheight]{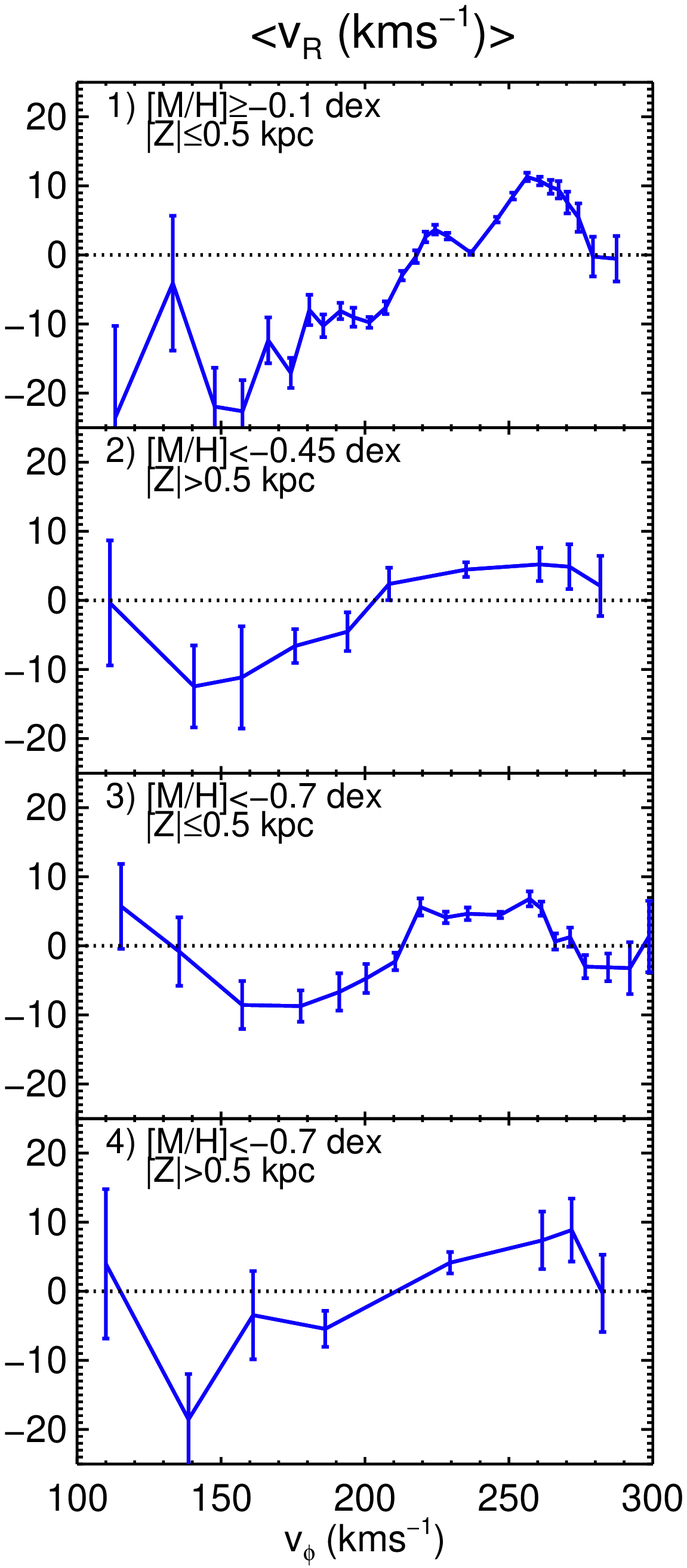}
\caption{Left: Difference in the number of counts between positive and negative $\vrr$, $\Delta=(N_+-N_-)$ (black curves) and the same number scaled to the total number of stars in both bins $\tilde\Delta=(N_+-N_-)/(N_++N_-)$ (red curves, right axis) as function $\vphi$ 
for the different subsets. Right: Mean $\vrr$ velocity for bins in $\vphi$ for the different samples.} 
\label{fig:dif1d}
\end{figure*}

To study this in more detail, we use the density field shown in 
the right panels of Fig.~\ref{fig:velocity-rave2} 
to compute the difference between regions with positive and negative $\vrr$. 
In practise, if
${\sigma_+}$ is the density in a certain pixel $(\vrr,\vphi)$  of the grid (of $2\kms$ size) 
and $\sigma_-$
is the number for the symmetric pixel $(-\vrr,\vphi)$, we compute 
$\Delta = \sigma_+-\sigma_-$.
This quantity is plotted in Fig.~\ref{fig:dif2d}. Red
colors indicate $\Delta > 0$ (more stars with $\vrr>0$), and blue colors $\Delta < 0$.
 
The upper left panel in Fig.~\ref{fig:dif2d}, corresponding to the
thin disk sample, clearly shows that the region with
 $\vphi\gtrsim240\kms$ has an excess of stars with $\vrr>0$, while the
contrary is true for $\vphi\lesssim240\kms$. This separation is not at
constant $\vphi$ for all $\vrr$ (i.e horizontal in the $\vrr$-$\vphi$
diagram) but rather varies with $\vrr$.  The other three panels,
corresponding to the thick disk subsets, depict a similar
asymmetry. However, in these cases, the asymmetry is not as sharp and
clear as for the thin disk set. This is probably due to the larger
velocity errors and to the lower number
of stars, especially for samples (2) and (4). Nevertheless, the asymmetry is very
clear for sample (3), and it is located roughly at the same velocities
as for sample (1).  Despite limitations for samples (2) and (4), it is
still clear that red colors dominate the half upper part of the
distribution while blue colors dominate in the other half
\footnote{Note that a wrong 
assumption of the peculiar velocity of the Sun $\Us$ would also produce 
an asymmetry in the counts of $\vrr>0$ with respect to $\vrr<0$, 
but this would be positive or negative everywhere and would not depend on $\vphi$ as we see here. Note also that a different assumption of the values for $\Ro$, $\Vs$ and $\vo$ can not produce the observed asymmetries, only shifting the positions or the velocity $\vphi$.}.

The left panel of Fig.~\ref{fig:dif1d} is a one-dimensional version of
Fig.~\ref{fig:dif2d}, where we plot for each $\vphi$ the difference
between the counts for $\vrr>0$ ($N_+$) and those for $\vrr<0$ ($N_-$). We use the method presented in
\citet{Scargle13} and implemented in \citet{VanderPlas12} to bin the
data in $\vphi$. It is a non-parametric technique that finds the
optimal data segments of variable size that maximize a certain fitness
function in a Bayesian likelihood framework and based on Poissonian
statistics.
Although the binning choice is arbitrary, we have checked that our conclusions do not change if we use bins of equal size or bins with equal number of stars. For this figure we plot also in red (right vertical axis) the relative asymmetry in the counts, i.e.  $\tilde{\Delta}=(N_+-N_-)/(N_++N_-)$. 
  
This plot shows the trend already highlighted in the two-dimensional
plots of Fig.~\ref{fig:dif2d}. There is a large asymmetry in the counts towards 
$\vrr>0$ that peaks at $\vphi\sim230-250\kms$ and extends from
$\vphi\sim200$ to $\sim275\kms$. A smaller but significant (note the
small errors) asymmetry $\Delta<0$ is
detected for velocities below $\sim200\kms$ and at least down to
$120-140\kms$. This asymmetry is in a region of the
velocity plane with a lower number of stars, and is thus better seen for the red
curves which corresponds to the normalized counts $\tilde\Delta$.

%
%

On the right panel of Fig.~\ref{fig:dif1d} we show the mean $\vrr$ as
function of $\vphi$ for all the subsets using the same binning as
before. The shape of these curves are similar to those on the left plot. This
is because an excess of counts for positive $\vrr$ reflects
in a positive mean $\vrr$, and conversely. The change in sign of the
mean $\vrr$ is significant for all four samples and occurs at 
$\vphi\sim200-220\kms$. It is noteworthy that the negative asymmetry 
seen both in the counts and in the mean $\vrr$ extends well into low $\vphi$,
a region more often thought  
to be characteristic of accretion events. For instance, this is
where the Arcturus stream is also located \citep{Navarro04}.

For the thin disk sample (top panels of Fig.~\ref{fig:dif1d})
we observe also other smaller bumps and more detailed features.
In particular, two clear positive peaks are seen 
 both in the counts $\tilde\Delta$ and in the mean $\vrr$. These are due to known
kinematic over-densities in the Solar Neighborhood. Some hints of
similar bumps (e.g. at $\vphi\sim215-220\kms$ and at $\vphi\sim250-260\kms$) are also seen
for the thick disk sample (3).

The results presented above are robust to the specific choice of proper motion catalog, as we have checked that no significant differences are seen when 
using other catalogs, namely UCAC2 and UCAC3 \citep[][and references therein]{Zacharias10}, Tycho-2 \citep{Hog00} and PPMXL \citep{Roeser10}. We also obtained the same results when we scaled the distances by factors of 0.8 and 1.2, i.e. assuming that the distances were overestimated and underestimated by 20\%, respectively. Furthermore the choice of the volume of the cylinder has no effect on our results: stars with the same cuts but located in a cylindrical ring of radius between 0.5 and $1.\kpc$ (dominated by giant stars instead of by dwarfs as in  our initial cylinder) show the same asymmetries.

As mentioned in Section \ref{sec:intro}, the main effect of the bar on
the local velocity distribution is a bimodality in the velocity
distribution near the Outer Lindblad Resonance \citep{Kalnajs91,Dehnen00}. This
produces the Hercules stream, an excess of stars with negative $\vrr$
at velocities around $\vphi=190\kms$ (heliocentric velocity
$V\sim-50\kms$), and the dominant mode of low-velocity stars centered
around the local standard of rest ($\vphi=240\kms$) and with an
elongation through positive $\vrr$. We believe that the asymmetries
that we observe have the same origin.




\section{Discussion and Conclusions}
\label{sec:concl} 

We have found clear asymmetries in the $\vrr$-$\vphi$ velocity
distributions of thin and thick disk stars near the Sun. In the thin disk such
asymmetries are due to well-known over-densities such as the 
Hercules stream, which has been explained by the resonant effects of the bar
near its OLR.  

This is the first time that the same type of structures and
asymmetries are seen in the thick disk. The features are significant
for the three different thick disk samples considered based on
metallicity and height above the plane, which we have estimated to
have low contamination from thin disk and halo stars. These findings suggest that the Galactic bar leaves strong
imprints on both the thin and thick disk.


 It is clear that the observed $\vphi$ velocities are highly skewed, not following a Gaussian distribution. \citet{Binney14} also showed that the radial and vertical velocities can not be fitted by Gaussian functions not only due to moving groups but also because they peak more sharply than Gaussians. Our results also imply that the velocities of thick disk stars should no
longer be described by independent Schwarzschild, multivariate
Gaussian or purely axisymmetric distributions: the reported asymmetry is both $\vphi$ and
$\vrr$ dependent. The RAVE dataset has also shown peculiar vertical velocity patterns \citep{Williams13} and other studies based on simulations have found that the non-axisymmetries of the Galaxy can also influence the vertical velocity distribution \citep{Faure14}. It follows that that dynamical models of the Galaxy
must fit the 3D velocity distributions, rather than the projected 1D. 

 
Simulations show that the deformations and over-densities of the
velocity distribution caused by the bar change with position (both in
radius and azimuthal angle) in the thin (e.g. \citealt{Dehnen00,Antoja14}) and
in the thick disk as well \citep{Monari13}. This implies that the
 asymmetries in the velocity distribution of the thick disk
must change with spatial position. Interestingly, rotational lags and asymmetries 
in the thick disk  were reported by \citet{Humphreys11} and \citet{Jayaraman13},  which may be further evidence of this. 
Future modeling of the velocity distributions must therefore be
position dependent. One should notice that the asymmetry projects differently on line of sight and transverse
velocity, creating different signatures. 

This also implies that the
nature of previously found substructures in the thick disk needs to be
revisited as these could be associated with dynamical resonances rather
than to accretion events. Specifically, the Arcturus stream would
seem to be naturally explained in this way, being an extension 
towards lower $\vphi$ velocities, 
which would also be favored given its chemical abundances \citep{Williams09},
and hence there would be no need to invoke
any accretion events 
nor ringing due to a past merger event \citep{Minchev09}. Also the substructure reported by
\citet{Gilmore02} could perhaps be explained along similar lines, although it was suggested that this is part of a metal-weak thick disk \citep{Kordopatis13c}.
The role of the bar on the formation of such structures should thus be investigated.

It is clear that with the advent of larger and more precise samples of the disk kinematics we are entering a new era where the classic velocity distribution models are not sufficient and the effects of the non-axisymmetry in the disk have to be taken into account in the modeling. This is particularly relevant now given that in approximately two years time the first Gaia data will be published and we expect to detect  these and perhaps other asymmetries far beyond the Sun.

\acknowledgments

TA is supported by an ESA Research Fellowship in Space Science. We thank Ralf Scholz for constructive comments on the analysis. AH was partially supported by ERC-StG
GALACTICA-240271. GM is supported by the {\it Centre National d'Etudes Spatiales} (CNES). Funding for RAVE has been provided by: the Anglo-Australian Observatory; the Leibniz-Institut fuer Astrophysik Potsdam; the Australian National University; the Australian Research Council; the French National Research Agency; the German Research foundation; the Istituto Nazionale di Astrofisica at Padova; The Johns Hopkins University; the National Science Foundation of the USA (AST-0908326); the W.M. Keck foundation; the Macquarie University; the Netherlands Research School for Astronomy; the Natural Sciences and Engineering Research Council of Canada; the Slovenian Research Agency; the Swiss National Science Foundation; the Science \& Technology Facilities Council of the UK; Opticon; Strasbourg Observatory; and the Universities of Groningen, Heidelberg and Sydney. 
The RAVE web site is at http://www.rave-survey.org. We thank the anonymous referee for the helpful comments.

\clearpage

\end{document}